\documentclass{aastex}

\newcommand{\emaila}{ E-mail: mfullana@mat.upv.es}
\newcommand{\emailb}{ E-mail: aalfonsofaus@yahoo.es}

\begin{document}

\title{Cosmic Background Bose Condensation (CBBC)
}

\shorttitle{CBBC}
\shortauthors{A. Alfonso-Faus and M. J. Fullana i Alfonso}

\author{Antonio Alfonso-Faus\altaffilmark{1}}
\affil{Escuela de Ingenier\'{\i}a Aeron\'autica y del Espacio, 
Plaza del Cardenal Cisneros, 3,
Madrid, 28040, Spain.\\
\emailb}
\and
\author{M\`arius Josep Fullana i Alfonso\altaffilmark{2}}
\affil{Institut de Matem\`atica Multidisciplin\`aria, 
Universitat Polit\`ecnica de Val\`encia, 
Cam\'{\i} de Vera, 
Val\`encia, 46022, Spain.\\
\emaila}

\begin{abstract}
Degeneracy effects for bosons are more important for smaller particle mass, smaller temperature and higher number density. 
Bose condensation requires that particles be in the same lowest energy quantum state. 
We propose a cosmic background Bose condensation, present everywhere, whith its particles having
the lowest quantum energy state, $\hbar c / \lambda$, with $\lambda$ about the size of the visible universe, and therefore unlocalized. 
This we identify with the quantum of the self gravitational potential energy of any particle, and with the bit of information of minimum energy. 
The entropy of the universe ($\sim 10^{122} \ bits$) has the highest number density 
($ \sim 10^{36} \ bits / cm^3$) of particles inside the visible universe, the smallest mass, $\sim 10^{-66} g$, and the smallest temperature, $\sim 10^{-29} K$. 
Therefore it is the best candidate for a  
Cosmic Background Bose Condensation (CBBC),
a completely calmed fluid, with no viscosity, in a superfluidity state, and possibly responsible for the expansion of the universe.
\end{abstract}


\section{Introduction}	

\cite{Wei} advanced a clue to suggest that large numbers are determined by both, microphysics and the influence of the whole universe. He constructed a mass using the physical constants $G, \hbar, c$ and the Hubble parameter $H$. This mass was not too different from the mass of a typical elementary particle (like a pion) and is given by

\begin{equation}
m \approx (\hbar^2 H /Gc)^{1/3}  
\label{e1}
\end{equation}

In our work here we consider a general elementary particle of mass $m$. This particle may include not only baryons but the possible quantum masses of dark matter and dark energy in the universe. Since the mass $m$ will disappear from the resultant relation, the conclusion is totally independent on the kind of elementary particle that we may consider. The self gravitational potential energy $E_g$ of this quantum of mass $m$ 
(and size its Compton wavelength $ \hbar /mc$) is given by  

\begin{equation}
E_g = G m^2 / (\hbar / mc) = G m^3 c / \hbar
\label{e2}
\end{equation}

This relation has been previously used in another context (\citeauthor{Siv} \citeyear{Siv}).  %, \citeyear{Siv2}). 
Combining  (\ref{e1}) and (\ref{e2}) we eliminate the mass $m$ to obtain

\begin{equation}
E_g \approx H \hbar
\label{e3}
\end{equation}

Here $\hbar$ is Planck's constant, usually interpreted as the smallest quantum of action (angular momentum). 
Since $H$ is of the order of $1/t$, $t$ the age of the universe ($t$ being a maximum time today), (\ref{e3}) is the lowest quantum energy state that it may exist. It is equivalent to $\hbar c / \lambda$ with $\lambda$ of the order of the size of the visible universe 
(it is the lowest quantum energy state with $\lambda \approx ct$). 
We identify it with the quantum of the self gravitational potential energy of any quantum particle (\citeauthor{AAFa} \citeyear{AAFa}). We also identify it with the bit, the unit of information with minimum energy (\citeauthor{AAF2} \citeyear{AAF2}). \cite{LLo}, about 10 years ago, stated that {\it Merely by existing, all physical systems register information} and about 25 years ago \cite{Lan}, as cited by \cite{LLo}, 
stated {\it Information is physical". And today we say here: All physical systems of mass M (energy $Mc^2$) are equivalent to an amount of information in number of bits of the order of }
\begin{equation}
Number \ of \ bits \ \approx M c^2 / E_g \approx Mc^2 / (H \hbar)
\label{e4}
\end{equation}

The equivalence between information and energy, as implied by the above relation, can be interpreted as the result of a recent experiment  
(\citeauthor{Fun} \citeyear{Fun}) where it is shown that entanglement can produce a gain in thermodynamic work, the gain being determined by a change of information content. Also a link between information theory and thermodynamics has been experimentally verified (\citeauthor{Ber} \citeyear{Ber}). Previously, in 2010,  an experimental demonstration of information-to-energy conversion was also published (\citeauthor{Toy} \citeyear{Toy}). 
Relation (\ref{e4}) has general, universal validity. In this sense the unit of energy, that should naturally be taken as the minimum quantum of energy 
$H \hbar$, implies that the relativistic energy of any mass $M$ has $N$ times this minimum quantum of energy, $N H \hbar$, being $N$ its number of information bits.
Therefore, $N H \hbar$ corresponds to the energy of all the information $N$ that carries the physical system. 
And as far as entropy $S$ is concerned, this number is also the same as $S/ (k_B \log 2)$, $k_B$ being Boltzmann constant, 
as we can talk about a generalized relation for entropy, in accordance with the ideas introduced by \cite{Lan2}:

\begin{equation}
S = N k_B \log 2 = \frac{Mc^2}{H \hbar} \ k_B \log 2
\label{e5}
\end{equation}

We see that the extensive property of entropy is preserved because, in accordance with our proposal, it comes to be proportional to the relativistic energy of the system. 
In this work we propose a Cosmic Background Bose condensation (CBBC), present everywhere, where the particle with minimum 
energy $E_g$ and mass $m_g = E_g /c^2$ is defined in (\ref{e3}). It is composed of very low energy and temperature components, with very high number density, and of course all in the same state.

\cite{Bek} found an upper bound for the ratio of the entropy $S_B$ to the energy $E = Mc^2$ of any bounded system with effective size $R$:
\begin{equation}
S_B / E < 2 \pi k_B R / \hbar c
\label{eq.1}
\end{equation}

About ten years later (\citeauthor{Hoo} \citeyear{Hoo}; \citeauthor{Sus} \citeyear{Sus}), a holographic principle was proposed giving a bound for the entropy $S_h$ of a bounded system of effective size $R$ as
\begin{equation}
S_h \leq  \pi k_B c^3 R^2 / \hbar G
\label{eq.2}
\end{equation}

The Bekenstein bound (\ref{eq.1}) is proportional to the product $MR$, while the holographic principle bound (\ref{eq.2}) 
is proportional to the area $R^2$. If the two bounds are identical (hence $M$ is proportional to $R$) 
here we prove that the system obeys the Schwarzschild condition for a black hole. 
We analyze this conclusion for the case of a universe with finite mass $M$ and a Hubble size $R \approx  ct$, 
$t$ being the age of the universe. $M$ and $R$ are obviously the maximum values they can have in our universe, the visible universe.

Also, for the case of a black hole we find that the Hawking and Unruh temperatures are the same. 
Then, for our universe we obtain the mass of the gravity quanta, of the order of $10^{-66} g$.

\section{Consequences of the identification of the two bounds}

Identifying the bounds (\ref{eq.1}) and (\ref{eq.2}) we get
\begin{equation}
2 M = c^2 R / G    
\label{eq.3}
\end{equation}
which is the condition for the system $(M, R)$ to be a black hole. Then, its entropy is given by the Hawking relation (\citeauthor{Haw} \citeyear{Haw})
\begin{equation}
S_H = \frac{4 \pi k_B}{\hbar c} \ G M^2
\label{eq.4}
\end{equation}
that coincides with the two bounds (\ref{eq.1}) and (\ref{eq.2}). The conclusions here are exclusively related to the Schwarzschild radius within the context of Einstein's general relativity.

The mass of the universe $M_u$ is a maximum. And so is its size $R$. A bounded system implies a finite value for both. 
Using present values for $M_u \approx 10^{56} g$ and $R \approx 10^{28} cm$ they fulfill the Schwarzschild condition (\ref{eq.3}). 
This is an evidence for the Universe to be a black hole (\citeauthor{AAF1} \citeyear{AAF1}). And its entropy today is about $10^{122} k_B \log 2$.

\section{The case for the Unruh and Hawking temperatures}

The fact that a black hole has a temperature, and therefore an entropy (\citeauthor{Haw} \citeyear{Haw})
implies that an observer at its surface, or event horizon, sees a perfect blackbody radiation, 
a thermal radiation with temperature $T_H$ given by
\begin{equation}
T_H =  \hbar c^3 / (8 \pi GMk) 
\label{eq.5}
\end{equation}
where $\hbar$ is the Planck's constant, $c$ the speed of light, $G$ the gravitational constant, $k$ Boltzmann constant 
and $M$ the mass of the black hole. This observer feels a surface gravitational acceleration $R''$. 
According to the \cite{Unr} effect an accelerated observer also sees a thermal radiation at a temperature $T_U$, 
proportional to the acceleration $R''$ and given by
\begin{equation}
T_U =  \hbar R'' / (2 \pi ck) 
\label{eq.6}
\end{equation}

Based upon the similarity between the mechanical and thermo dynamical properties of both effects, (\ref{eq.5}) 
and (\ref{eq.6}), we identify both temperatures and find the relation:
\begin{equation}
R'' =  c^4 / (4GM) 
\label{eq.7}
\end{equation}

Identifying the Unruh acceleration to the surface gravitational acceleration: 
\begin{equation}
R'' =  GM/R^2
\label{eq.8}
\end{equation}
and substituting in (\ref{eq.7})  we finally get
\begin{equation}
2GM/c^2 = R
\label{eq.9}
\end{equation}

This is the condition for a black hole. Since the Hawking temperature refers to a black hole 
this result confirms the validity of the identification of the two temperatures, (\ref{eq.5}) and (\ref{eq.6}), 
as well as the interpretation of the Unruh acceleration in (\ref{eq.8}).

\section{Application of the cosmological principle}

The cosmological principle may be stated with the two special conditions of the universe: it is homogeneous and isotropic. 
This means that, on the average, all places in the universe are equivalent (at the same {\it time}) 
and that observing the universe at one location it looks the same in any direction. 
The cosmic microwave background radiation is a good example, a blackbody radiation at about $2.7 K$. 
This implies that there is no center of the universe, or equivalently, that any local place is a center. 

We have seen that the universe may be taken as a black hole, and therefore that it makes sense to think that there must be an event horizon, a two dimensional bounding surface around each observer. If all places in the universe are equivalent then all places can say this, and the natural event horizon is the Hubble sphere, with radius $R$ about $c / H \approx 10^{28} cm$ today. Then at any point in the universe it can be interpreted as a two dimensional spherical surface, may be in a {\it virtual} sense. Following the holographic principle, all the information of the three dimensional world, as we see it, is contained in this spherical surface. And any observer, following the cosmological principle, can be seen as being at the center of the three dimensional {\it sphere}. To combine both principles, the cosmological and the holographic, we can think of an isotropic, spherically symmetric, acceleration present at each point in the universe given by 
(\ref{eq.8}). This is a change of view from a 3 dimensional one to a 2 dimensional world. Also we have an isotropic temperature given by
\begin{equation}
T =  \hbar c^3 / (8 \pi GMk)  = (1/4 \pi k) 1 / R \approx  10^{-29} K
\label{eq.10}
\end{equation}

This is the temperature of the gravitational quanta (\citeauthor{AAF2} \citeyear{AAF2}) at the present time. The equivalent mass of one quantum of gravitational potential energy 
is then from (\ref{eq.10}) found to be about $10^{-66} g$. This may be interpreted as the ultimate quantum of mass. 
Its wavelength (in the Compton sense) is of the order of the size of the universe, $ct \approx 10^{28} cm$ and therefore it is a gravity quanta unlocalized in the universe, as the gravitational field
(\citeauthor{Mis} \citeyear{Mis}).  It is a boson and not a photon, the photon being the quantum of the electromagnetic field that should be localized in the universe.
 
The scale factor between the Planck scale and our universe today is about $10^{61}$. 
Multiplying the temperature found in (\ref{eq.10}) by this numerical scale factor 
we get the Planck's temperature $T_p \approx 10^{32} K$ at the Planck's time $10^{-44} s$, 
when the universe had the Planck's size $l_p  \approx 10^{-33} cm$.       

\section{Cosmic Background Bose Condensation (CBBC)}
We now present the conditions favorable to have a Bose condensation as derived from the energy (\citeauthor{Eis} \citeyear{Eis}) per particle relation $E/N$: 

\begin{equation}
E/N = \frac{3}{2} \ kT \left \{ 1 - \frac{1}{2^{5/2}} \ \frac{N h^3}{V (2 \pi m k_B T)^{3/2}} \right \}
\label{e5}
\end{equation}

The term beyond 1 in the above bracket is the deviation of the Bose gas from the classical gas, the degeneracy effect. 
As we can see in the formulation (\ref{e5}) this effect for bosons is more important the smaller the particle mass is ( $\sim 10^{-66} g$ in our case), 
the smaller the temperature is ( $\sim 10^{-29} K$ in our case) and the higher the number density is ($\sim 10^{36} \ particles/cm^3$ ). 
Then, the particle we are presenting here appears to be a good candidate for a universal Bose condensation background. 

\section{The CBBC and the expansion of the universe}
As we have seen in (\ref{e3}) the gravity quanta proposed here does not depend on the related origin: baryons, dark matter or dark energy. 
Usually the expansion of the universe is considered to be related to the cosmological constant $\Lambda$. This constant was introduced by Einstein to avoid the collapse of the universe due to attractive gravitation by means of an outward pressure given by $\Lambda$. Within this context we know today that the percentages of baryonic content in the universe, dark matter and dark energy, are respectively about 4\%, 27\% and 69\%. This is given in terms of the usual dimensionless $\Omega$:
\begin{equation}
\Omega_b \approx 0.04, \ \ \ \ \ \Omega_{DM} \approx 0.27, \ \ \ \ \ \Omega_{\Lambda} \approx 0.69
\label{e6}
\end{equation}
\noindent
all of them adding up to 1. Given that our approach here does not discriminate between baryons, dark matter and dark energy, we may attribute the CBBC expansion effect as due to its pressure as related to the baryon component and possibly dark matter quanta or even dark energy quanta. We do not know today if there are quanta in the dark components. The maximum effect would correspond to consider that the total, critical density, is responsible for the gravity quanta presented here as the ground component of the CBBC, and then its effect would correspond to $\Omega_{CBBC} = 1$.

\section{Conclusions}

The universe can be seen as a black hole. From each observer in it we can interpret that he/she is at the center of a sphere, with the Hubble radius. 
From this we interpret the spherical surface as the event horizon, a two dimensional surface that follows the physics of the holographic principle.

The isotropic acceleration present at each point in the universe, and given by (\ref{eq.8}), 
implies that there is no distortion for the spherically distributed acceleration, as imposed by the cosmological principle. 
However, the presence of a nearby important mass, like the sun, will distort this spherically symmetric picture. 
With respect to the probes Pioneer 10/11, that detected an anomalous extra acceleration towards 
the sun of value ($8.74 \pm 1.33) \times 10^{-8} cm/s^2$ (\citeauthor{And} \citeyear{And}), 
we can see that this value is only a bit higher than the one predicted by (\ref{eq.8}), 
which is about $7.7 \times 10^{-8}cm/s^2$. 
This difference is an effect that can be explained by the presence of the sun converting the isotropic acceleration to an anisotropic one.

Similarly there may be a factor, due to the influence of nearby masses (i.e. a massive black hole at the center of the galaxy), 
in the cases of the observed rotation curves in spiral galaxies. They imply that the speed of stars, 
instead of decreasing with the distance $r$ from the galactic center, 
is constant or even increases slowly when far from the central luminous object (\citeauthor{Dre} \citeyear{Dre}). For the case of globular clusters (\citeauthor{Sca} \citeyear{Sca}), 
where no dark matter is expected to be present, we have stronger evidence in support of the existence of the acceleration field. 
Also the escape velocity at the sun location, with respect to our galaxy, is higher than expected. 
The earth-moon distance increases with time and there is a residual part not explained by tidal effects. 
And the same occurs for the planets in the solar system. We present this evidence in support of the universal field of acceleration, $R''$ (\citeauthor{AAF3} \citeyear{AAF3}).

Finally, for the case of a black hole we find that the Hawking and Unruh temperatures are the same. For our universe we obtain the mass of the gravity quanta, of the order of $10^{-66} g$, with a wavelength that corresponds to its size. It may be identified with the information bit, with entropy $k$. A Cosmic Background Bose Condensation fits well with the properties found here for the quantum of gravitational potential energy.


\begin{thebibliography}{}

\bibitem[Alfonso-Faus(2010a)]{AAFa} Alfonso-Faus, A., 2010a, "Universality of the self gravitational potential energy of any fundamental particle", Astrophysics and Space Science, 337, 363

\bibitem[Alfonso-Faus(2010b)]{AAF1} Alfonso-Faus, A., 2010b,  "The case for the Universe to be a quantum black hole", Astrophysics and Space Science, 325, 113

\bibitem[Alfonso-Faus(2010c)]{AAF3} Alfonso-Faus, A., 2010c,  "Galaxies: kinematics as a proof of the existence of a universal field of minimum acceleration", preprint, arXiv:0708.0308

\bibitem[Alfonso-Faus(2011)]{AAF2}Alfonso-Faus, A., 2011, "Quantum gravity and information theories linked by the physical properties of the bit", 
preprint, arXiv: 1105.3143 

\bibitem[Anderson et al.(1998)]{And}  Anderson, J. D., et al., 1998, "Indication, from Pioneer 10/11, Galileo, and Ulysses Data, of an Apparent Anomalous, Weak, Long-Range Acceleration", Physical Review Letters, 81, 2858

\bibitem[Bekenstein(1981)]{Bek} Bekenstein, J. D., 1981, Physical Review D, 23 (2), 287

\bibitem[B\'erut et al.(2012)]{Ber} B\'erut, A. et al., 2012, "Experimental verification of Landauer's principle linking information and thermodynamics", Nature, 483, 187

\bibitem[Drees \& Chung-Lin(2007)]{Dre} Drees, M., \& Chung-Lin, S.,  2007, "Theoretical Interpretation of Experimental Data from Direct Dark Matter Detection", Journal of Cosmology and Astroparticle Physics, 0706, 011

\bibitem[Eisberg \& Resnick(1985)]{Eis} Eisberg, R. and Resnick, R., 1985, "Quantum Physics of atoms, molecules, solids, nuclei and particles" 1985, Second Edition, John Wiley \& Sons Inc., New York

\bibitem[Funo, Watanabe \& Ueda(2012)]{Fun} Funo, K., Watanabe, Y. \& Ueda, M., 2012, "Thermodynamic Work Gain from Entanglement", preprint, quant-ph/1207.6872

\bibitem[Hawking(1974)]{Haw} Hawking, S. W., 1974, "Black hole explosions?" , Nature, 248, 30

\bibitem[Landauer(1961)]{Lan2} Landauer, R., 1961, "Irreversibility and Heat Generation in the Computing Process",
IBM Journal of Research and Development, 5, 183

\bibitem[Landauer(1988)]{Lan} Landauer, R., 1988, "Dissipation and noise immunity in computation and communication", Nature, 335, 779

\bibitem[Lloyd(2002)]{LLo} Lloyd, S., 2002, "Computational capacity of the universe", Physical Review Letters 88, 237901 

\bibitem[Misner, Thorne \& Wheeler(1973)]{Mis} Misner, C.W., Thorne, K.S. \& Wheeler, J.A., 1973, "Gravitation", W.H. Freeman and Company, Reading (England), p.466 ("Why the energy of the gravitational field cannot be localized")

\bibitem[Scarpa \& Falomo(2010)]{Sca} Scarpa, R. \& Falomo, R., 2010, "Testing Newtonian gravity in the low acceleration regime with globular clusters: the case of omega Centauri revisited", Astronomy and Astrophysics, 523, A43

\bibitem[Sivaram(1982)]{Siv} Sivaram, C., 1982, "Cosmological and quantum constraint on particle masses", American Journal of Physics, 50, 279

%\bibitem[Sivaram(1999)]{Siv2} Sivaram, C., 1999, "Alternative to Dark Matter and some implications for the Early Universe", 
%in "The Early Universe", ed V. B. Johri, Hadronic Press, Molise, Italia, p. 33

\bibitem[Susskind(1995)]{Sus} Susskind, L., 1995, "The World as a hologram", Journal of Mathematical Physics, 36, 6377

\bibitem['t Hooft(1993)]{Hoo} 't Hooft, G., 1993, "Dimensional Reduction in Quantum Gravity", preprint, arXiv:gr-qc/9310026

\bibitem[Toyabe et al.(2010)]{Toy} Toyabe, S. et al., 2010, "Experimental demonstration of information-to-energy conversion and validation of the generalized Jarzynski equality", 
Nature Physics, 6, 988

\bibitem[Unruh(1976)]{Unr} Unruh, W.G., 1976, "Notes on black-hole evaporation", Physical Review D, 14 (4), 870

\bibitem[Weinberg(1972)]{Wei} Weinberg, S.,  1972, "Gravitation and Cosmology: Principles and applications of the General Theory of Relativity", John Wiley \& Sons Inc., New York, p.619


\end{thebibliography}
\end{document}